\documentclass[12pt]{elsart}
\usepackage{feynmp,epsfig,graphics}
\def\be{\begin{eqnarray}}
\def\ee{\end{eqnarray}}
\def\bc{\begin{center}}
\def\ec{\end{center}}

\newcommand{\tr}{\rm tr \,}

\begin{document}
GSI-Preprint-2003-23
\begin{frontmatter}
\title{On charm baryon resonances and chiral symmetry }
\author[GSI,TU]{M.F.M. Lutz}
\author[NBI]{and E.E. Kolomeitsev}
\address[GSI]{Gesellschaft f\"ur Schwerionenforschung (GSI)\\
Planck Str. 1, 64291 Darmstadt, Germany}
\address[TU]{Institut f\"ur Kernphysik, TU Darmstadt\\
D-64289 Darmstadt, Germany}
\address[NBI]{The Niels Bohr Institute\\ Blegdamsvej 17, DK-2100 Copenhagen, Denmark}
\begin{abstract}
We study heavy-light baryon resonances with quantum numbers $J^P\!=\!\frac{1}{2}^-$  in terms
of the non-linear chiral SU(3) Lagrangian. Within the $\chi-$BS(3) approach
a parameter-free leading order prediction is obtained  for the scattering of Goldstone
bosons off heavy-light baryon resonances with $J^P=\frac{1}{2}^+$. The three
states $\Lambda_{c1}(2593), \Lambda_{c0}(2880)$
and $\Xi_{c1}(2790)$ discovered by the CLEO collaboration are recovered. We suggest the
existence of resonance states that form an anti-quindecimplet, two sextet and two
anti-triplet representations of the SU(3) group. In particular, narrow states with
anomalous isospin (I) and strangeness (S) quantum numbers
$(I,S)=(\frac{1}{2},+1)$ are anticipated.
\end{abstract}
\end{frontmatter}

\section{Introduction}

In  a recent work \cite{KL03} it was demonstrated that chiral
coupled-channel dynamics generates heavy-light meson resonances
with quantum numbers $J^P\!=\!0^+$ and $J^P\!=\!1^+$. Such states
were first predicted in \cite{NRZ93,BH94} based on the chiral
quark model and recently observed for the first time by the BABAR
and CLEO collaborations \cite{BaBar,CLEO}. Due to the different
dynamical assumptions of the chiral quark model versus the chiral
coupled-channel approach the theoretical prediction are quite
different (see \cite{BEH03,NRZ03} and references therein). Most
spectacular is the prediction of $J^P\!=\!0^+$ and $J^P\!=\!1^+$
heavy-light meson states with negative strangeness. Whereas the
chiral quark model implies an anti-triplet of open-charm or
open-bottom states only, an additional sextet of states was
predicted in \cite{KL03}. These findings suggest that the chiral
SU(3) symmetry plays  a decisive role also in the physics of
heavy-light baryons. Double-charm or double-bottom baryon states
are completely analogous to the heavy-light mesons and all results
from \cite{KL03} can be straightforwardly taken over to this
sector. Therefore in this paper we focus on baryons with one charm
quark. The chiral quark model predict $\frac{1}{2}^+$ and
$\frac{3}{2}^+$ states that form anti-triplet and sextet
representations of the SU(3) group, respectively
\cite{NRZ93,BH94,BEH03,NRZ03}. The main goal of this work is to
unravel the consequences of chiral coupled-channel dynamics for
such states. Recently three $J^P=\frac{1}{2}^-$ states were
observed by the CLEO Collaboration \cite{Edwards,Csorna,Artuso}.
Open charm baryons have been studied extensively in the literature
\cite{Isgur,Neubert,Bigi,Pirjol,Chiladze,Korner1,Korner2,LLS00,ChCG01,BChCG01,Blechman}.
For review articles on the heavy-quark effective theory approach
we refer to \cite{Neubert,Bigi}.

We apply the $\chi$-BS(3) approach developed originally for meson-baryon scattering
\cite{LK00,LK01,LH02,LK02,Granada,Copenhagen} that is based on the {\bf chiral} SU({\bf{3}})
Lagrangian and formulated in terms of the {\bf{B}}ethe-{\bf{S}}alpeter equation.
The latter was applied recently successfully also to meson-meson scattering \cite{LK03,KL03}.
Using the chiral SU(3) Lagrangian with heavy-light baryon fields that transform
non-linearly under the chiral SU(3) group a coupled-channel description of the meson-baryon
scattering in the open charm sector is developed. The crucial importance of
coupled-channel dynamics for the baryon-resonance formation in the u-,d-,s-sector of QCD
was first anticipated in a series of works in the sixties
\cite{Wyld,Dalitz,Ball,Rajasekaran,Logan}. Related works
based on the chiral SU(3) Lagrangian are \cite{weise,OR,OM,grnpi,grkl,Oset-prl,Oset-plb,Jido03}.
Our major result is the prediction that there exist open-charm states
with $J^P\!=\!\frac{1}{2}^-$ quantum numbers forming one anti-quindecimplet,
two anti-triplet and two sextet representations of the SU(3) group. This differs from the results
implied by the linear realization of the chiral SU(3) symmetry leading to
one anti-triplet and one sextet only. We recover the so far known
$J^P\!=\!\frac{1}{2}^-$ resonance states established by the CLEO
collaboration \cite{Edwards,Csorna,Artuso}. Most spectacular is the
promise of new $J^P\!=\!\frac{1}{2}^-$ states with anomalous quantum numbers
$(I,S)=(\frac{1}{2},+1),(\frac{3}{2},-1),(1,-2)$.

\section{Chiral coupled-channel dynamics: the $\chi$-BS(3) approach}

The starting point to study the scattering of Goldstone bosons off
heavy-light baryons is the chiral SU(3) Lagrangian.
We identify the leading order interaction Lagrangian
density \cite{Wein-Tomo,Wise92,YCCLLY92,BD92,Cho94,Cheng97} describing the interaction of
Goldstone bosons with heavy-light baryons,
\begin{eqnarray}
{\mathcal L}(x) &=&
\frac{i}{16\,f^2}\,\tr  \Big(
\bar H_{[\bar 3]}(x)\,\gamma^\mu\, \Big[ H_{[\bar 3]}(x) \,,[\phi (x) , (\partial_\mu\,\phi(x))]_-\Big]_+ \Big)
\nonumber\\
&+&\frac{i}{16\,f^2}\,\tr  \Big(
\bar H_{[6]}(x)\,\gamma^\mu\, \Big[ H_{[6]}(x) \,,[\phi (x) , (\partial_\mu\,\phi(x))]_-\Big]_+ \Big)
  \,,
\label{WT-term}
\end{eqnarray}
with the Goldstone bosons field $\phi$ and
massive baryon fields $H_{[\bar 3]}$ and $H_{[6]}$. The Weinberg-Tomozawa term (\ref{WT-term})
follows by gauging the kinetic term of the heavy-baryon fields with the chiral SU(3) group
and expanding the resulting expression in powers of the Goldstone bosons fields.
The parameter $f$ in (\ref{WT-term}) characterizes the associated covariant derivative and
is known from the weak decay process of the
pions. We use $f= 90$ MeV through out this work.
Since we will assume perfect isospin symmetry it is convenient to decompose
the fields into their isospin multiplets. The fields can be written in terms of
isospin multiplet fields like $K =(K^{(+)},K^{(0)})^t $ and $\Xi_c=(\Xi^{(+)}_c,\Xi_c^{(0)})^t$,

\begin{table}[t]
\tabcolsep=2mm
\begin{tabular}{|cccccccccccc|}
\hline\hline
\multicolumn{12}{|c|}{$(I,S)_{[\bar 3]}$}\\
 \hline\hline \multicolumn{3}{|c|}{$(\frac{1}{2},+1)$} &
\multicolumn{3}{c|}{\phantom{xxxxxx}$(0,0)$\phantom{xxxxxx}} & \multicolumn{3}{c|}{$(1,0)$} &
\multicolumn{3}{c|}{$(\frac12,-1)$}
\\ \hline
\multicolumn{3}{|c|}{$(\Lambda_c\,K)$} &
\multicolumn{3}{c|}{$\left(\begin{array}{c} (\Lambda_c\,\eta )\\
({\textstyle{1\over \sqrt{2}}}\,K^t \, i\,\sigma_2\, \Xi_c)
\end{array}\right)$}&
\multicolumn{3}{c|}{$\left(\begin{array}{c} (\Lambda_c \, \pi)\\ (
{\textstyle{1\over \sqrt{2}}}\,K^t \, i\,\sigma_2\,\sigma\, \Xi_c)
\end{array}\right)$} &
\multicolumn{3}{c|}{$\left(\begin{array}{c} ({\textstyle{1\over
\sqrt{3}}}\,\pi \cdot \sigma\, \Xi_c)\\
(\Lambda_c\,i\,\sigma_2\,\bar K^t) \\ ( \eta \,\Xi_c)
\end{array}\right)$}
\\\hline
\multicolumn{4}{|c|}{\phantom{xxxxxx}$(\frac32,-1)$\phantom{xxxxxx}}
& \multicolumn{4}{c|}{\phantom{xxxxxx}$(0,-2)$\phantom{xxxxxx}} &
\multicolumn{4}{c|}{$(1,-2)$}
\\\hline
\multicolumn{4}{|c|}{$( \pi \cdot T\,\Xi_c)$} &
\multicolumn{4}{c|}{$( {\textstyle{1\over \sqrt{2}}}\,\bar K
\,\Xi_c)$} & \multicolumn{4}{c|}{$({\textstyle{1\over
\sqrt{2}}}\,\bar K \,\sigma\,\Xi_c)$}
\\\hline\hline
\multicolumn{12}{|c|}{$(I,S)_{[6]}$}\\
 \hline\hline
\multicolumn{3}{|c|}{$(\frac{1}{2},+1)$} &
\multicolumn{3}{c|}{$(\frac{3}{2},+1)$} &
\multicolumn{3}{c|}{$(0,0)$} &
\multicolumn{3}{c|}{$(1,0)$} \\
\hline
\multicolumn{3}{|c|}{$\frac{1}{\sqrt 3}\,(\Sigma_c\,\sigma\,K)$} &
\multicolumn{3}{|c|}{$(\Sigma_c\cdot T\,K)$} &
\multicolumn{3}{c|}{$\left(\begin{array}{c}
\frac{1}{\sqrt{3}}\,(\Sigma_c\cdot \pi )\\
\frac{1}{\sqrt{2}}\,(K^t \, i\,\sigma_2\, \Xi'_c)
\end{array}\right)$} &
\multicolumn{3}{c|}{$\left(\begin{array}{c}
\frac{i}{\sqrt{2}}\,(\Sigma_c\times \pi )\\
( \eta \,\Sigma_c) \\
\frac{1}{\sqrt{2}}\,(K^t \, i\,\sigma_2\,\sigma\, \Xi'_c)
\end{array}\right)$}\\
\hline
\multicolumn{3}{|c|}{$(2,0)$} &
\multicolumn{3}{c|}{$(\frac12,-1)$} &
\multicolumn{3}{c|}{$(\frac32,-1)$} &
\multicolumn{3}{c|}{$(0,-2)$} \\
\hline
\multicolumn{3}{|c|}{$(\Sigma_c\cdot S\cdot \pi )$} &
\multicolumn{3}{c|}{$\left(\begin{array}{c}
\frac{1}{\sqrt  3}\,(\pi \cdot \sigma\,\Sigma_c )\\
(\Sigma_c \, \eta)\\
(\Omega_c\,K)
\end{array}\right)$} &
\multicolumn{3}{c|}{$
\left(\begin{array}{c}( \pi \cdot T\,\Xi'_c) \\ (\Sigma_c\cdot T\,i\,\sigma_2\,\bar K^t)
\end{array}\right)$} &
\multicolumn{3}{c|}{$
\left(\begin{array}{c}\frac{1}{\sqrt 2}\,( \bar K\,\Xi'_c) \\ (\Omega_c\,\eta)
\end{array}\right)$}
\\ \hline
\multicolumn{6}{|c|}{$(1,-2)$} &
\multicolumn{6}{c|}{$(\frac{1}{2},-3)$}
\\ \hline
\multicolumn{6}{|c|}{$
\left(\begin{array}{c} \Omega_c\,\pi \\
\frac{1}{\sqrt 2}\,(\bar K \,\sigma\,\Xi'_c)
\end{array}\right)$} &
\multicolumn{6}{|c|}{$(\Omega_c\,i\,\sigma_2\,\bar K^t)$}
\\\hline\hline
\end{tabular}
\caption{The definition of coupled-channel states with  $(I,S)_{[\bar 3]}$ and
$(I,S)_{[6]}$. Here $\sigma=(\sigma_1,\sigma_2,\sigma_3)$ are the isospin Pauli
matrices. The isospin transition operator $T$ connects isospin-$\frac12$ and
isospin-$\frac32$ states. It is normalized by $T^\dag_i\, T_j=\delta_{ij}-\sigma_i\, \sigma_j/3$.
The matrix valued vector $S_{[n]}$ couples two isospin-1 states into a spin-2 state. It
satisfies $\sum_{n=1}^5 S^\dag_{[n],ac}\, S_{[n],bd}=\frac12\, \delta_{ab}\,\delta_{cd}+
\frac12\, \delta_{ad}\, \delta_{cb}-\frac13\, \delta_{ac}\,\delta_{bd}$.}
\label{tab:states}
\end{table}

\begin{eqnarray}
&& \phi = \tau \cdot \pi (140)
+ \alpha^\dagger \cdot  K (494) +  K^\dagger(494) \cdot \alpha
+ \eta(547)\,\lambda_8\,,
\nonumber\\
&& H_{[\bar 3]}  = {\textstyle{1\over \sqrt{2}}}\,\alpha^\dagger \cdot \Xi_c(2470)
- {\textstyle{1\over \sqrt{2}}}\,\Xi_c^t(2470)\cdot \alpha
+  i\,\tau_2\,\Lambda_c(2284) \,,
\nonumber\\
&& H_{[6]} = {\textstyle{1\over \sqrt{2}}}\,\alpha^\dagger \cdot \Xi'_c(2580)
+ {\textstyle{1\over \sqrt{2}}}\,\Xi_c^{'t}(2580)\cdot \alpha
+ \Sigma_c(2453) \cdot \tau \,i\,\tau_2
\nonumber\\
&& \qquad \quad+  {\textstyle{\sqrt{2}\over 3}}\, \big(1-\sqrt{3}\,\lambda_8 \big)\,\Omega_c(2704)  \,,
\nonumber\\
&& \nonumber\\
&& \alpha^\dagger = {\textstyle{1\over \sqrt{2}}}\,(\lambda_4+i\,\lambda_5 ,\lambda_6+i\,\lambda_7 )\,,\qquad
\tau = (\lambda_1,\lambda_2, \lambda_3)\,,
\end{eqnarray}
where the matrices $\lambda_i$ are the standard Gell-Mann generators of the SU(3) algebra.
The numbers in the brackets recall the approximate masses of the particles in units of
MeV \cite{PDG02}.

\begin{table}[t]
\tabcolsep=4.4mm
\begin{center}
\begin{tabular}{||c||p{9.0mm}|p{9.0mm}|p{9.0mm}|p{9.0mm}|p{9.0mm}|p{9.0mm}|p{9.0mm}|p{9.0mm}|p{9.0mm}||}
\hline
$(I, S)_{[\bar 3]}$    & ($\frac{1}{2}, +1$)  & ($0, 0$) & ($1, 0$) &
($\frac{1}{2},-1$) & ($\frac{3}{2},-1$) & ($0,-2$) &($1,-2$)  \\
 \hline\hline
11 & $-1$ & $\phantom{-}0$ & $\phantom{-}0$& $\phantom{-}2$ & $-1$ & $\phantom{-}1$ & $-1$  \\ 
  \hline
12 &-- & $-\sqrt{3}$ &$\phantom{-}1$ & $\phantom{-}\sqrt{\frac{3}{2}}$ &-- & -- & --  \\ 
  \hline
22 &-- & $\phantom{-}2$& $\phantom{-}0$  & $\phantom{-}1$& -- & -- & --  \\ 
  \hline
13 &-- & -- &-- & $\phantom{-}0$ &-- & -- & --  \\ 
  \hline
23 &-- & -- &-- & $-\sqrt{\frac{3}{2}}$ &-- & -- &-- \\ 
  \hline
33 &-- &-- &-- &$\phantom{-}0$ &-- &--  &--  \\ 
\hline
\end{tabular}
\caption{The coefficients $C^{(I,S)}$ that characterize the  interaction of
Goldstone bosons with the heavy baryon fields $H_{[\bar 3]}$ as introduced in (\ref{lwt}). The
ordering of the states is introduced in Tab. \ref{tab:states}}
\label{tab:coeff-3}
\end{center}
\end{table}

The scattering problem decouples into ten orthogonal
channels specified by isospin (I) and strangeness (S) quantum numbers,
\begin{eqnarray}
&& (I,S)_{[\bar 3]}= (({\textstyle{1\over{2}}},+1), (0,0), (1, 0), ({\textstyle{3\over{2}}}, -1),
({\textstyle{3\over{2}}}, -1), (0,-2), (1,-2))\,,
\nonumber\\
&& (I,S)_{[6]}-(I,S)_{[\bar 3]}=
( ({\textstyle{3\over{2}}},+1),(2,0), ({\textstyle{1\over{2}}}, -3))\,,
\label{lwt}
\end{eqnarray}
where the scattering of Goldstone bosons off the anti-triplet leads to seven channels but
the scattering off the sextet to additional three channels. At leading order the two
sectors $(I,S)_{[\bar 3]}$ and
$(I,S)_{[6]}$ do not couple to each other. Only subleading terms in the
chiral Lagrangian lead to processes like $\pi \,\Xi_c \to K\,\Omega_c$.
In Tab. \ref{tab:states} the channels that contribute in a given sector $(I,S)$ are listed.
Heavy-light baryon resonances with quantum numbers $J^P\!=\!\frac{1}{2}^-$
manifest themselves as poles in the s-wave scattering amplitudes, $M^{(I,S)}_{[\bar 3]}(\sqrt{s}\,)$,
and, $M^{(I,S)}_{[6]}(\sqrt{s}\,)$, which in the
$\chi-$BS(3) approach \cite{LK02,LK03} take the simple form
\begin{eqnarray}
&&  M^{(I,S)}(\sqrt{s}\,) = \Big[ 1- V^{(I,S)}(\sqrt{s}\,)\,J^{(I,S)}(\sqrt{s}\,)\Big]^{-1}\,
V^{(I,S)}(\sqrt{s}\,)\,.
\label{final-t}
\end{eqnarray}

\begin{table}[t]
\tabcolsep=2.4mm
\begin{center}
\begin{tabular}{||c||p{7mm}|p{7mm}|p{7mm}|p{7mm}|p{7mm}|p{7mm}|p{7mm}|p{7mm}|p{7mm}|p{7mm}||}
\hline
$(I, S)_{[6]}$            &11 &12 &22 &13 &23 &33 &14 &24 &34 &44 \\
 \hline\hline
($\frac12$, $+1$) &$\phantom{-}1$ &-- &-- &-- &-- &-- &-- &-- &-- & -- \\ 
  \hline
($\frac32$, $+1$) &$-1$ &-- & -- &-- &-- &-- &-- &-- &-- & -- \\ 
  \hline
(0,0) & $\phantom{-}4$ & $\phantom{-}\sqrt{3}$ & $\phantom{-}2$ &-- &-- &-- &-- &-- &-- & -- \\ 
  \hline
(1,0) &$\phantom{-}2$ &$\phantom{-}0$ &$\phantom{-}0$  &$\phantom{-}\sqrt{2}$ &$-\sqrt{3}$ &$\phantom{-}0$ & --& --& --& -- \\ 
  \hline
(2,0) &$\phantom{-}0$ & -- & -- &-- &-- &-- &-- &-- &-- & -- \\ 
  \hline
($\frac12$, $-1$) &$\phantom{-}2$ &$-\frac{1}{\sqrt 2}$ &$\phantom{-}3$ & $\phantom{-}0$ & $-\frac{3}{\sqrt{2}}$ & $\phantom{-}0$ & $\phantom{-}\sqrt{3}$ & $\phantom{-}0$& $\phantom{-}\sqrt{3}$ & $\phantom{-}2$ \\ 
  \hline
($\frac32$, $-1$) &$-1$ &$\phantom{-}\sqrt{2}$ &$\phantom{-}0$ &-- &-- &-- &-- &-- &-- &-- \\ 
  \hline
(0, $-2$) &$\phantom{-}1$ &$-\sqrt 6$ &$\phantom{-}0$ &-- &-- &-- &-- &-- &-- & -- \\ 
  \hline
(1, $-2$) &$\phantom{-}0$ &$-\sqrt{2}$ &$-1$ &-- &-- &-- &-- &-- &-- & -- \\ 
  \hline
($\frac12$, $-3$) & $-2$ &--&-- &-- & -- & -- &-- &-- &-- & -- \\ 
\hline
\end{tabular}
\caption{The coefficients $C^{(I,S)}$ that characterize the  interaction of
Goldstone bosons with heavy baryon fields $H_{[6]}$ as introduced in (\ref{lwt}).
The ordering of the
states is introduced in Tab. \ref{tab:states}}
\label{tab:coeff-6}
\end{center}
\end{table}
The effective interaction kernel $V^{(I,S)}(\sqrt{s}\,)$ in (\ref{final-t}) is determined by
the leading order chiral SU(3) Lagrangian (\ref{WT-term}),
\begin{eqnarray}
V^{(I,S)}(\sqrt{s}\,) = \frac{C^{(I,S)}}{4\,f^2}\, \Big(
2\,\sqrt{s}-M-\bar M \Big) \,,
\label{VWT}
\end{eqnarray}
where $M$ and $\bar M$ are the masses of initial and final baryon states. We use
capital $M$ for the masses of heavy-light baryons and small $m$ for the masses of the Goldstone
bosons. The matrix of coefficients $C^{(I,S)}$ that characterize the interaction strength in
a given channel is given in Tab.~\ref{tab:coeff-3} and \ref{tab:coeff-6}.
The s-wave interaction kernels are identical for the two scattering problems considered here.
The loop functions, diagonal in the coupled-channel space, are
\begin{eqnarray}
&& J(\sqrt{s}\,) =
\Big(M + (M^2+p_{\rm cm}^2)^{1/2} \Big)\,
\Big(I(\sqrt{s}\,)-I(\mu) \Big)\,,
\nonumber\\
\nonumber\\
&& I(\sqrt{s}\,)=\frac{1}{16\,\pi^2}
\left( \frac{p_{\rm cm}}{\sqrt{s}}\,
\left( \ln \left(1-\frac{s-2\,p_{\rm cm}\,\sqrt{s}}{m^2+M^2} \right)
-\ln \left(1-\frac{s+2\,p_{\rm cm}\sqrt{s}}{m^2+M^2} \right)\right)
\right.
\nonumber\\
&&\qquad \qquad + \left.
\left(\frac{1}{2}\,\frac{m^2+M^2}{m^2-M^2}
-\frac{m^2-M^2}{2\,s}
\right)
\,\ln \left( \frac{m^2}{M^2}\right) +1 \right)+I(0)\;,
\label{i-def}
\end{eqnarray}
where $\sqrt{s}= \sqrt{M^2+p_{\rm cm}^2}+ \sqrt{m^2+p_{\rm cm}^2}$.
A crucial ingredient of the $\chi-$BS(3) scheme is its approximate crossing
symmetry guaranteed by a proper choice of the subtraction scales $\mu_{[\bar 3]}^{(I,S)}$
and $\mu_{[6]}^{(I,S)}$.
Referring to the detailed discussions
in \cite{LK02,Granada,Copenhagen,LK03} we define
\begin{eqnarray}
&& \mu_{[\bar 3]}^{(I,0)} =  M_{\Lambda_c(2284)}\,,\quad
\mu_{[\bar 3]}^{(I, \pm 1)} = M_{\Xi_c(2470)}\,,\quad
\mu_{[\bar 3]}^{(I, -2)} = M_{\Lambda_c(2284)} \,,
\nonumber\\
&& \mu_{[6]}^{(I,0)} = M_{\Sigma_c(2453)}\,,\quad
\mu_{[6]}^{(I, \pm 1)} = M_{\Xi'_c(2580)}\,,\quad
\mu_{[6]}^{(I, -2)} = M_{\Omega_c(2704)} \,,
\nonumber\\
&& \mu_{[6]}^{(I, -3)} = M_{\Xi'_c(2580)}  \,.
\label{mu-def}
\end{eqnarray}
Given the subtraction scales (\ref{mu-def}) the leading order calculation presented in this work
is parameter free. Of course chiral correction terms do lead to further so
far unknown parameters  which need to be adjusted to data. Within the $\chi-$BS(3)
approach such correction terms enter the effective interaction kernel $V$ rather than leading to
subtraction scales different from (\ref{mu-def}). In particular the leading correction effects
are determined by the counter terms of chiral order $Q^2$.
The effect of altering the subtraction scales away from their optimal values (\ref{mu-def})
can be compensated for by incorporating counter terms in the chiral Lagrangian that carry
order $Q^3$. Our scheme  has the advantage that once the
parameters describing subleading effects are determined in a subset of sectors one has
immediate predictions for all sectors (I,S). In order to estimate the size of correction
terms it is nevertheless useful to vary the subtraction scales around their optimal values.
With (\ref{final-t}-\ref{mu-def}) the brief exposition of the $\chi-$BS(3)
approach as  applied to heavy-light baryon resonances is completed.

\begin{figure}[t]
\begin{center}
\includegraphics[clip=true,width=13cm]{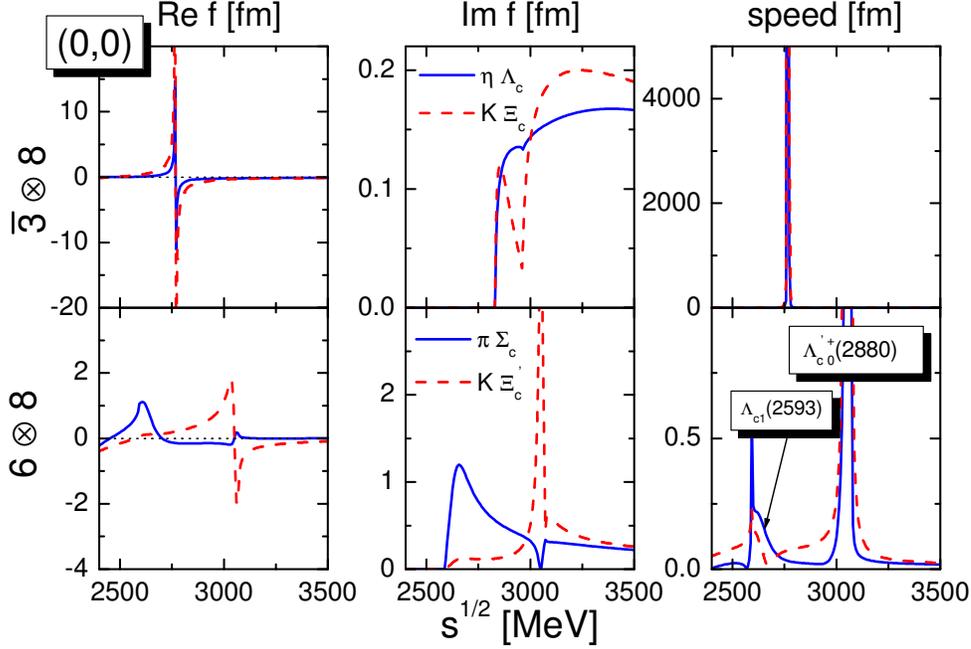}
\end{center}
\caption{Open charm baryon resonances with $J^P=\frac{1}{2}^-$  and $(I,S)=(0,0)$ as seen in
the scattering of Goldstone bosons off anti-triplet $(\Lambda_c(2284), \Xi_c(2470))$ and
sextet $(\Sigma_c(2453),\Xi'_c(2580),\Omega_c(2704))$ baryons.
Shown are speed plots together with real and imaginary parts of reduced scattering amplitude,
$f_{ab}$, with $t_{ab}= f_{ab}\,(p^{(a)}_{\rm cm}\,p^{(b)}_{\rm cm})^{1/2}$ (see (\ref{def-speed})).}
\label{fig:1}
\end{figure}

\begin{figure}[t]
\begin{center}
\includegraphics[clip=true,width=13cm]{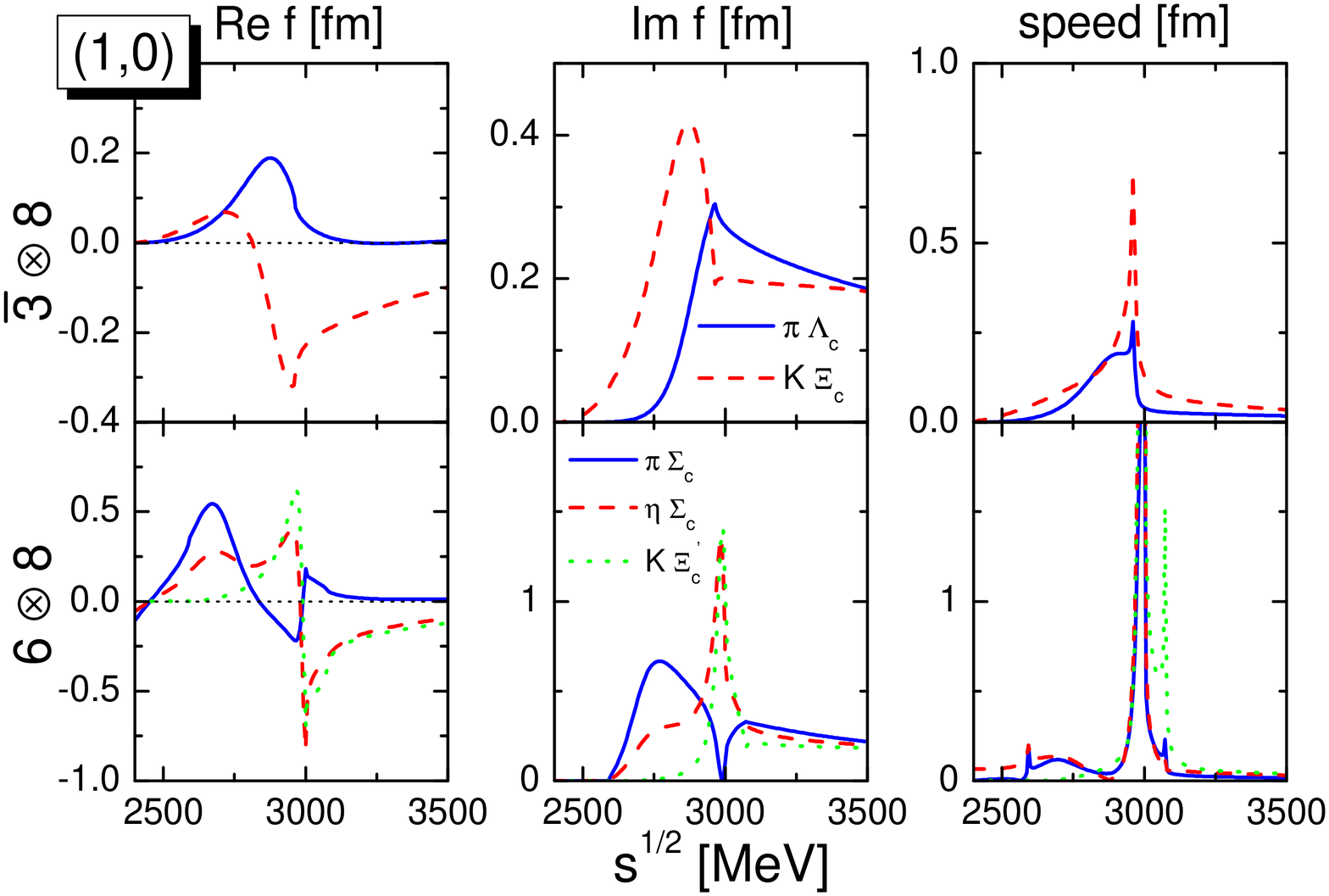}
\end{center}
\caption{Open charm baryon resonances with $J^P=\frac{1}{2}^-$  and $(I,S)=(1,0)$ as seen in
the scattering of Goldstone bosons off anti-triplet $(\Lambda_c(2284), \Xi_c(2470))$ and
sextet $(\Sigma_c(2453),\Xi'_c(2580),\Omega_c(2704))$ baryons.
Shown are speed plots together with real and imaginary parts of reduced scattering amplitude,
$f_{ab}$, with $t_{ab}= f_{ab}\,(p^{(a)}_{\rm cm}\,p^{(b)}_{\rm cm})^{1/2}$ (see (\ref{def-speed})).}
\label{fig:2}
\end{figure}

\begin{figure}[t]
\begin{center}
\includegraphics[clip=true,width=13cm]{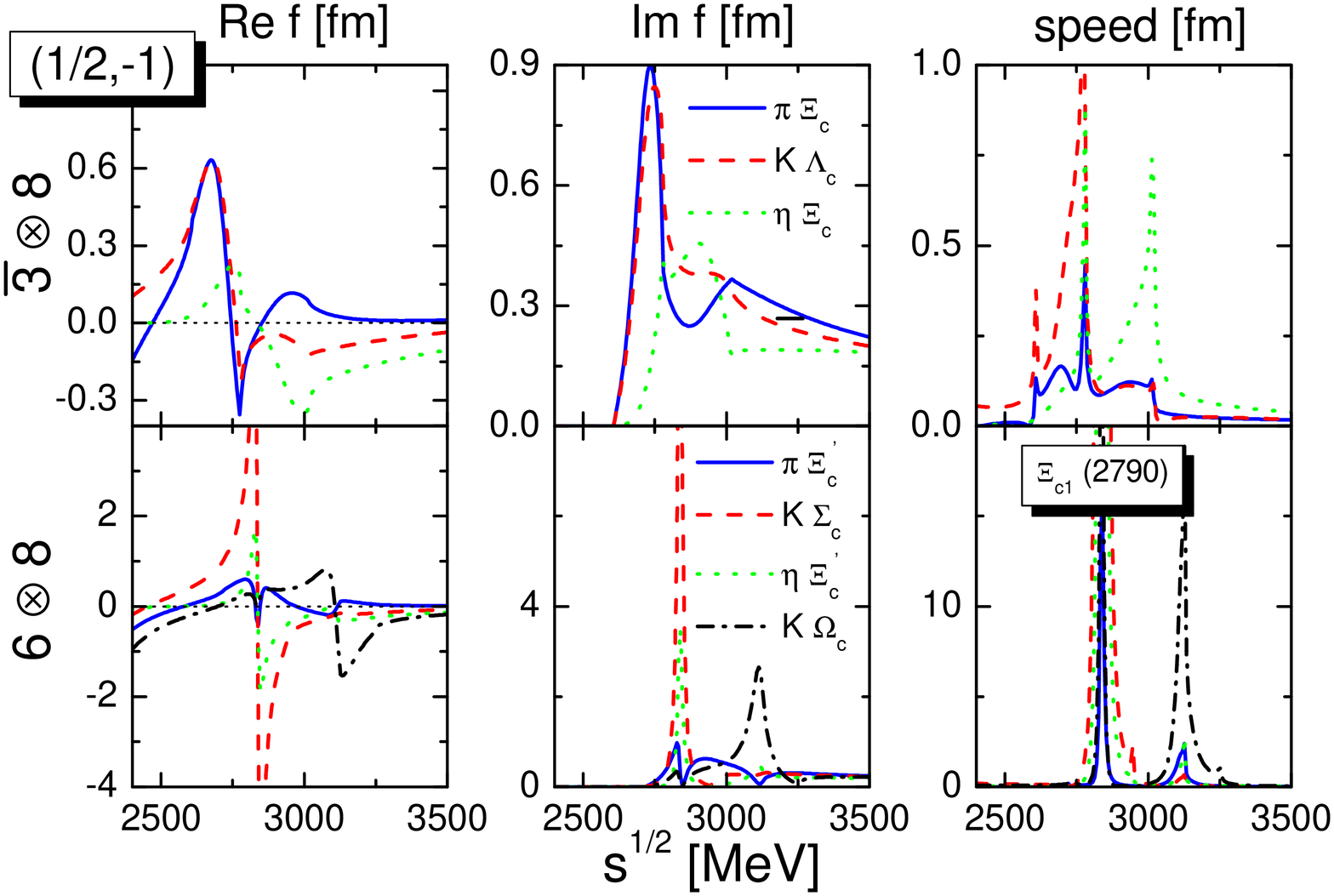}
\end{center}
\caption{Open charm baryon resonances with $J^P=\frac{1}{2}^-$  and $(I,S)=(\frac{1}{2},-1)$ as seen in
the scattering of Goldstone bosons off anti-triplet $(\Lambda_c(2284), \Xi_c(2470))$ and
sextet $(\Sigma_c(2453),\Xi'_c(2580),\Omega_c(2704))$ baryons.
Shown are speed plots together with real and imaginary parts of reduced scattering amplitude,
$f_{ab}$, with $t_{ab}= f_{ab}\,(p^{(a)}_{\rm cm}\,p^{(b)}_{\rm cm})^{1/2}$ (see (\ref{def-speed})).}
\label{fig:3}
\end{figure}

\begin{figure}[t]
\begin{center}
\includegraphics[clip=true,width=13cm]{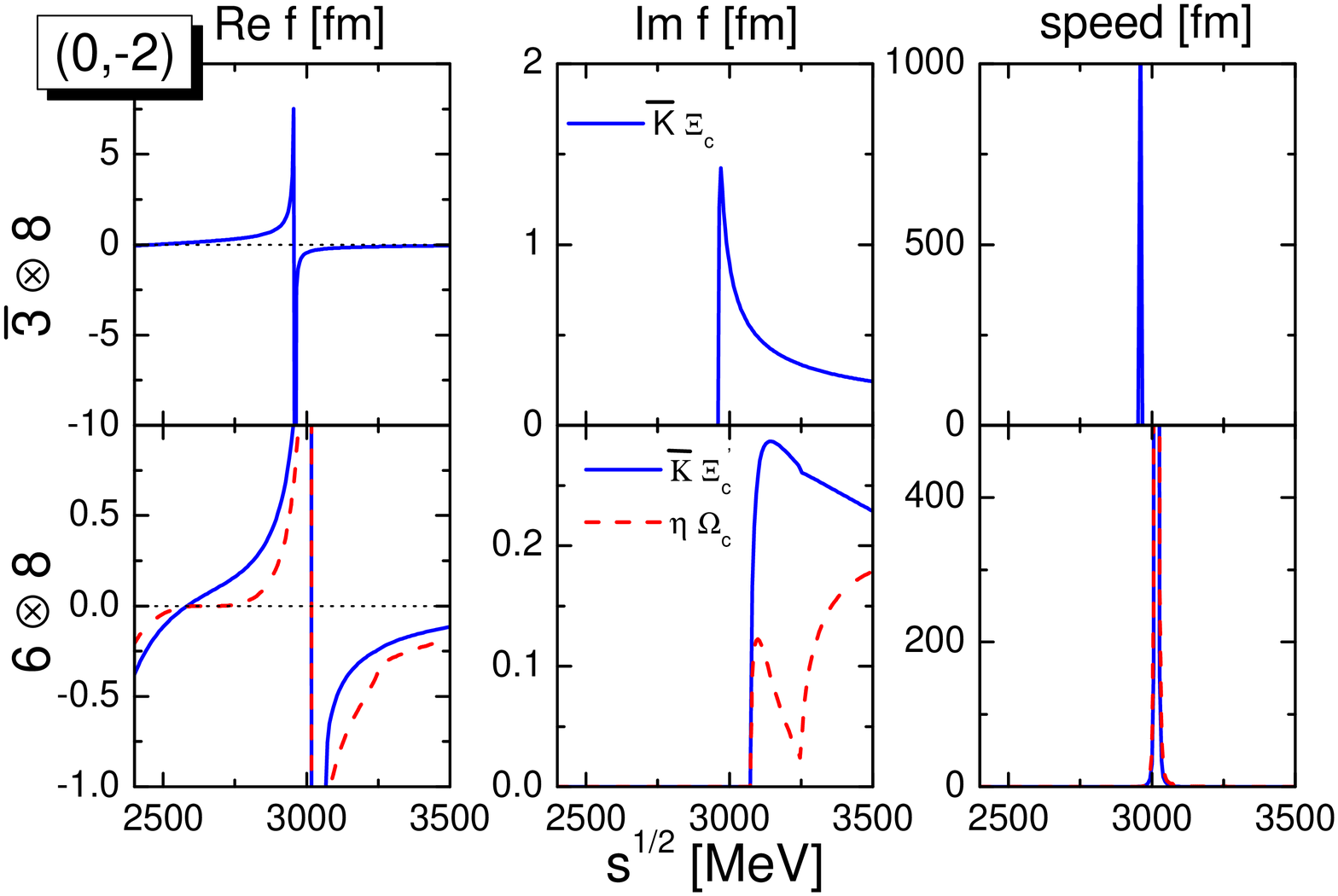}
\end{center}
\caption{Open charm baryon resonances with $J^P=\frac{1}{2}^-$  and $(I,S)=(0,-2)$ as seen in
the scattering of Goldstone bosons off anti-triplet $(\Lambda_c(2284), \Xi_c(2470))$ and
sextet $(\Sigma_c(2453),\Xi'_c(2580),\Omega_c(2704))$ baryons.
Shown are speed plots together with real and imaginary parts of reduced scattering amplitude,
$f_{ab}$, with $t_{ab}= f_{ab}\,(p^{(a)}_{\rm cm}\,p^{(b)}_{\rm cm})^{1/2}$ (see (\ref{def-speed})).}
\label{fig:4}
\end{figure}

\begin{figure}[t]
\begin{center}
\includegraphics[clip=true,width=13cm]{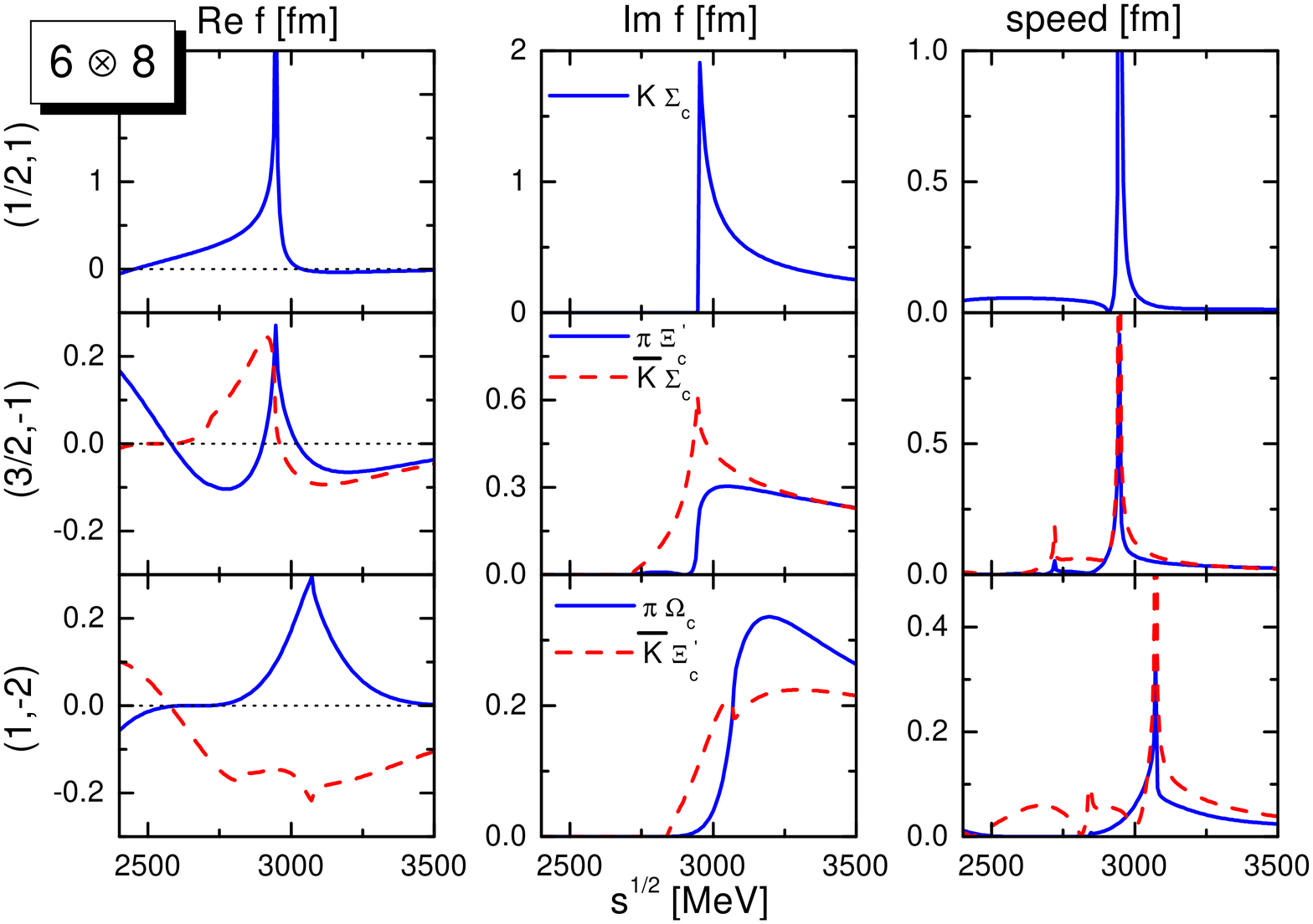}
\end{center}
\caption{Open charm baryon resonances with $(I,S)=(\frac{1}{2},1),(\frac{3}{2},-1),(1,-2)$
and $J^P=\frac{1}{2}^-$  as seen in
the scattering of Goldstone bosons off sextet
$(\Sigma_c(2453),\Xi_c(2580),\Omega_c(2704))$ baryons.
Shown are speed plots together with real and imaginary parts of reduced scattering amplitude,
$f_{ab}$, with $t_{ab}= f_{ab}\,(p^{(a)}_{\rm cm}\,p^{(b)}_{\rm cm})^{1/2}$ (see (\ref{def-speed})).}
\label{fig:5}
\end{figure}

\section{Results}

To study the formation of heavy-light baryon resonances we generate speed plots as suggested
by H\"ohler \cite{Hoehler:speed}. The speed ${\rm Speed}_{ab}^{(I,S)}(\sqrt{s})$ of a given
channel $a\,b$ is introduced by \cite{Hoehler:speed,speed},
\begin{eqnarray}
&& t^{(I,S)}_{ab}(\sqrt{s}\,)=\frac{1}{8\,\pi \,\sqrt{s}}\,
\Big((M_a+E_a)\, p^{(a)}_{\rm cm}\,(M_b+E_b)\,p_{\rm cm}^{(b)}\Big)^{1/2}\,M_{ab}^{(I,S)}(\sqrt{s}\,)\,,
\nonumber\\
&& {\rm Speed}_{ab}^{(I,S)}(\sqrt{s}\,) = \Big|\sum_{c}\,
 \Big[\frac{d}{d \,\sqrt{s}}\, t_{ac}^{(I,S)}(\sqrt{s}\,)\Big]\,
 \Big(\delta_{cb}+2\,i\,t_{cb}^{(I,S)}(\sqrt{s}\,) \Big)^\dagger
\Big| \,,
\label{def-speed}
\end{eqnarray}
where $E_a =(M_a^2+(p_{\rm cm}^{(a)})^2)^{1/2}$. If a resonance is formed in a scattering
process its $\rm{Speed}(\sqrt{s})$ will show a maximum at the resonance mass
(see e.g. \cite{LK03}).

In order to explore the SU(3) multiplet structure of  the resonance states
we first study the 'heavy' SU(3) limit \cite{Copenhagen,LK03,KL03}
with $m_{\pi ,K,\eta} = 500$ MeV. For the anti-triplet $J^P=\frac{1}{2}^+$ states we use
a somewhat arbitrary common mass $M=2400$ MeV. In this case we obtain an anti-triplet of
mass 2778 MeV with poles in the $(0,0),(\frac{1}{2},-1)$ amplitudes and a sextet
of mass 2900 MeV with poles in the $(1,0),(\frac{1}{2},-1),(0,-2)$ amplitudes.
The result is reasonably stable against small variations of the optimal subtraction scale.
Lowering the latter by 200 MeV reduces the anti-triplet and sextet masses by 40 MeV
and 5 MeV only. Our finding reflects that the Weinberg-Tomozawa interaction (\ref{WT-term}),
\begin{eqnarray}
\bar 3\otimes 8= \bar 3\oplus 6 \oplus \overline{15}
\label{}
\end{eqnarray}
is attractive in the anti-triplet and sextet channels but repulsive in the
anti-quindecimplet channel.
Similarly, using a common mass
for the sextet $J^P=\frac{1}{2}^+$ states of 2500 MeV we obtain
an anti-triplet of mass 2807 MeV with
poles in the $(0,0),(\frac{1}{2},-1)$ amplitudes, a sextet
of mass 2875 MeV with poles in the $(1,0),(\frac{1}{2},-1),(0,-2)$ amplitudes and an
anti-quindecimplet of mass 3000 MeV with poles in the
$((\frac{1}{2},1),(0,0),(1,0),(\frac{1}{2},-1),(\frac{3}{2},-1),(1,-2))$ amplitudes.
In this case the Weinberg-Tomozawa interaction (\ref{WT-term}),
\begin{eqnarray}
6 \otimes 8= \bar 3\oplus 6 \oplus \overline{15} \oplus 24
\end{eqnarray}
predicts attraction in the anti-triplet, sextet and anti-quindecimplet channels
but repulsion in the 24-plet channel.

In Figs. \ref{fig:1}-\ref{fig:4} the spectrum as it is predicted by the
$\chi-$BS(3) approach in terms of physical masses and the pion-decay constant
$f=90$ MeV is shown. In the upper (lower) panels the figures show the amplitudes
and speed plots describing the scattering of Goldstone bosons off the anti-triplet (sextet)
$J^P=\frac{1}{2}^+$ open charm baryons. At leading order in the chiral expansion
channels involving the anti-triplet and sextet open charm baryons decouple.
We predict a bound state of mass 2767 MeV in the $(0,0)_{[\bar 3]}$-sector
(see Fig. \ref{fig:1}). This state should be identified with the $\Lambda_c(2880)$
recently detected by the CLEO collaboration \cite{Artuso} via its decay into the
$\pi \Sigma_c(2453)\to \Lambda_c\pi\pi$ channel. The narrow width of the observed state
of smaller than 8 MeV \cite{Artuso} appears consistent with a suppressed coupling of that
state to the $\pi \Sigma_c(2453)$ channel as predicted by chiral symmetry. In the lower panel
of the figure the $(0,0)_{[6]}$-sector is presented. A resonance at about 2650 MeV
that couples strongly to the $\pi \Sigma_c(2453)$ channel is predicted. The properties of
this state are close to the ones of the $\Lambda_c(2593)$ resonance \cite{Edwards}.
Given the fact that our computation is parameter-free this is a remarkable result.
However, here we obtain a decay width which is significantly larger than the empirical width of
about 4 MeV \cite{Edwards}. Chiral correction terms that couple the states seen in the lower
and upper panels of Fig. \ref{fig:1} are expected to decrease this width. Level-level
repulsion of the two observed states should lower the mass of the lighter state but
push up the mass of the heavier state. A clear prediction of chiral-coupled channel dynamics is an
additional narrow resonance state in the $(0,0)_{[6]}$-sector at 3050 MeV as
part of the anti-quindecimplet discussed above.

In Fig. \ref{fig:2} our predictions for excitations of the $\Sigma_c$ baryon with
$J^P=\frac{1}{2}^-$ are displayed.  Two broad sextet states at about 2800 MeV that are
eventually expected to mix and one narrow state at 2985 MeV part of the anti-quindecimplet
are obtained. So far none of these states has been observed. In particular the latter
state which couples strongly to the $\eta \,\Sigma_c$ and $K\,\Xi_c'$ channels may
warrant a dedicated search for. In Fig. \ref{fig:3} the spectrum of the
$\Xi_c$ baryons with $J^P=\frac{1}{2}^-$ is shown. All together we expect
five states. In the $(\frac{1}{2},1)_{[\bar 3]}$-sector a clear resonance
that couples strongly to the $\pi\,\Xi_c $ and $K\,\Lambda_c$ channels
is seen at about 2750 MeV. At somewhat higher masses a weak resonance signal
is suggested in the $\eta \,\Xi_c$ channel. In the lower panel of Fig. \ref{fig:3}
two narrow states at 2830 MeV and 3120 MeV are shown. The lower state couples strongly
to the $K\,\Sigma_c$ and $\eta \,\Xi_c$ channels and due to its small width should be
identified with the $\Xi_c(2790)$ resonance \cite{Csorna}. The latter state has an empirical
width smaller than 15 MeV \cite{Csorna}. The upper state has a strong coupling constant
to the $K\,\Omega_c$ channel. One would expect the width of these states to increase
somewhat once the subleading decay channels are incorporated. We do not find any
clear signal of a third state in the lower panel of Fig. \ref{fig:3}.
We turn to Fig. \ref{fig:4} which probes with $(0,-2)$ exclusively the sextet states. Two
distinct narrow states at 2959 MeV and 3016 MeV are anticipated. Finally in Fig. \ref{fig:5}
our results for the channels $(\frac{1}{2},1),(\frac{3}{2},-1),(1,-2)$ are displayed.
Only members of the anti-quindecimplet manifest themselves as states in these channels. A narrow
structure at 2947 MeV is predicted in the $(\frac{1}{2},1)$ channel. This state has a mass
that is very close to the $K\,\Sigma_c$-threshold. Therefore in the speed plot of
Fig. \ref{fig:5} it is difficult to separate the resonance signal from the square-root
singularity induced by the opening of the corresponding channel. In the remaining
channels $(\frac{3}{2},-1),(1,-2)$
no clear resonance signal is seen. The bound-state obtained in the 'heavy' SU(3) limit
become rather broad resonances at around 3000 MeV.

{\bfseries{Acknowledgments}}

The authors acknowledge stimulating discussions with M.A. Nowak.

\end{document}